\documentclass[twocolumn,a4paper,floatfix,twoside,showpacs,prl]{revtex4}

\usepackage{graphicx}
\usepackage{amsmath}
\usepackage{amssymb}
\usepackage{bm}
\usepackage{units}
\usepackage{wasysym}

\newcommand{\im}{i}

\newcommand{\version}{July 16, 2008}

\DeclareMathOperator{\Real}{Re}

\renewcommand{\section}[1]{{\par\it #1.---}}

\begin{document}
\title{Qubit coherence decay down to threshold: influence of substrate dimensions}

\author{Roland Doll}
\email[E-mail:${\;\;}$]{roland.doll@physik.uni-augsburg.de}
\author{Peter H\"anggi}
\author{Sigmund Kohler}
\affiliation{%
Institut f{\"u}r Physik, Universit{\"a}t Augsburg, Universit\"atsstra{\ss}e
1, D-86135 Augsburg, Germany}
\author{Martijn Wubs}
\affiliation{%
Niels Bohr International Academy \& QUANTOP, The Niels Bohr Institute, Copenhagen University, DK 2100, Denmark }

\date{\version}

\begin{abstract}
Keeping single-qubit quantum coherence above some threshold value not far below
unity is a prerequisite for fault-tolerant quantum error correction (QEC). We
study the initial dephasing of solid-state qubits in the independent-boson
model, which describes well recent experiments on quantum dot (QD) excitons both
in bulk and in substrates of reduced geometry such as nanotubes. Using explicit
expressions for the exact coherence dynamics, a minimal QEC rate is identified
in terms of the error threshold, temperature, and qubit-environment coupling
strength. This allows us to systematically study the benefit of a current trend
towards substrates with reduced dimensions.
\end{abstract}


\maketitle

Two-level systems with long-lived quantum coherence are candidate qubits, the
units of quantum information \cite{Nielsen2000a}. Interactions between
solid-state qubits and substrate phonons cause decoherence. It seems natural to
fight decoherence by reducing the number of substrate degrees of freedom, by
going from bulk to planar or linear geometries. In fact, QDs can nowadays be
embedded in confined structures such as freestanding semiconductor membranes
\cite{Hoehberger2003,Weig2004} or nanotubes and nanowires
\cite{Sapmaz2006,Bjork2004}. Such substrates may allow tailoring of the phonon
spectrum and, thus, controlling the qubit dephasing. However, it is not obvious
whether fewer substrate dimensions do mean less decoherence. It is interesting
to compare photoluminescence measurements of single QDs in bulk
environment~\cite{Borri2001a,Krummheuer2002a,Borri2005a,Krummheuer2005a} and in
nanotubes~\cite{Hoegele2007_short,Galland2008_short}. In both cases, pure
dephasing due to deformation-potential coupling to acoustic phonons is the
dominant decoherence mechanism~\cite{Duke1965a,Takagahara1993a} since relaxation
occurs on a much longer time scale. Using a Markovian master equation, {\em
i.e.}~approximating the dephasing as exponential decay with a coherence time
$T_{2}$, one finds $T_{2}=\infty$ for bulk and a finite $T_{2}$ time for 1D
substrates (details below); this would be an argument {\em against} reducing
substrate dimensions. However, both in bulk and in reduced geometries, fast
dephasing at short times has been observed as a broad background in
spectra~\cite{Borri2001a,Krummheuer2002a,Borri2005a,Krummheuer2005a,Galland2008_short,Lindwall2007}.
According to~\cite{Galland2008_short}, it is this {\em non}-exponential decay
that may hamper applications for quantum information processing (QIP) with 1D
substrates. But are these really less ideal?

Future QIP devices will require built-in QEC, as some decoherence is inevitable.
From the information-theoretical side come stringent requirements for
fault-tolerant QEC: gate error levels $\epsilon$ should be less than $10^{-3}$
\cite{Aliferis2008a}, a value that may be relaxed in the future. Usually
assumptions such as local and Markovian decoherence  go into the derivation of
$\epsilon$. One may criticize these~\cite{Alicki2006a,Doll} and go beyond
them~\cite{Terhal2005a}. We start from the other end, with a given  $\epsilon$
and a realistic qubit-bath model.

In this paper, we study the minimal rate $\omega_\text{qec}$ at which
single-qubit errors should be corrected, requiring the coherence to stay above
the threshold value $1-\epsilon$. The beautiful experiments on QDs in bulk
(3D)~\cite{Borri2001a,Vagov2004a,Borri2005a} and   on nanotube excitons
(1D)~\cite{Galland2008_short} are both well explained by the so-called
independent-boson model, and we employ its generic version. We present
analytical expressions for $\omega_{\rm qec}$ and for the exact coherence
dynamics, with the substrate dimension as a free parameter, that reproduce the
measured rich dynamics and lead to new predictions: how can parameters best be
changed so that error correction is needed less frequently.

\section{Model}
To account for the coupling of the QD to quantized lattice
vibrations we follow Refs.~\cite{Krummheuer2002a,Mahan1990} and employ the
independent-boson model with Hamiltonian $H = H_0 + H_\text{qb}$ where $H_0 =
\Delta \sigma_z / 2 + \sum_\mathbf k \hbar\omega_{\mathbf k} b_\mathbf k^\dagger
b_\mathbf k$ describes the uncoupled system of QD and bosons, with Pauli matrix
$\sigma_z$ for the qubit and creation and annihilation operators $b_{\mathbf
k}^{(\dagger)}$ for phonons of mode $\mathbf k$. We focus on the dominant
deformation-potential coupling to acoustic phonons with dispersion
$\omega_\mathbf k = v_s |\mathbf k|$ where $v_s$ denotes the sound velocity. The
qubit-boson interaction $H_{\rm qb} = \hbar \sigma_z \sum_\mathbf k g_\mathbf k
(b_\mathbf k + b^\dagger_{-\mathbf k})$ is characterized by microscopic
couplings $g_\mathbf k \propto D_s \sqrt{|\mathbf k |} / \sqrt{2 \hbar \rho V
v_s}$ in terms of a deformation potential $D_s$ for the excitons, the mass
density $\rho$ and the volume
$V$~\cite{Takagahara1999a,Krummheuer2002a,Lindwall2007}. The more specialized
model in Refs.~\cite{Krummheuer2002a,Galland2008_short} takes different
deformation potentials and confinement lengths for electrons and holes into
account. As shown below, our simpler model is accurate and has the major
advantage that explicit expressions for the coherence decay can be obtained. The
phonon bath gives rise to the spectral density $J(\omega) = \sum_\mathbf k
|g_\mathbf k|^2 \delta (\omega - v_s |\mathbf k|)$ which for
deformation-potential coupling and spherical QD symmetry with confinement length
$l_s$ has the form $J_{s}(\omega) = \alpha_s \omega^s \omega_\text c^{1-s}
\exp(-\omega/ \omega_\text c)$ with cutoff frequency $\omega_\text c = v_s /
l_s$ and dimensionless coupling strength~$\alpha_s$.

\section{Exact dynamics for arbitrary dimension}
We find the explicit expression for the qubit coherence $c_s(t)$, {\em i.e.}~the
exciton polarization after a $\delta$-like excitation pulse. In terms of the
scaled temperature $\theta = k_\text B T / \hbar \omega_\text c$,  the dynamics
is given by $c_s(t) = \exp[-\lambda_s(t)]$, with  exponent
\begin{equation} \label{eqnCoherenceExact}
\begin{split}
\lambda_s(t) &= 8\alpha_s(-\theta)^{s-1} \left[ F^{(s-1)}(\theta) - \Real
F^{(s-1)}(\theta[1+ \im \omega_\text c t]) \right]\\
&\quad + 4\alpha_s\Gamma(s-1)\left(
\frac{\cos[(s-1)\arctan(\omega_\text c t)]}{(1+\omega^2_\text c t^2)^{(s-1)/2}} -1 \right),
\end{split}
\end{equation}
where $\Gamma(z)$ denotes Euler's Gamma function, $F(z) = \log \Gamma(z)$ and
its $n$-th derivative $F^{(n)}$ which for positive $n$ equals the Polygamma
function $\Psi_{n-1}(z)$. Hereby we generalize previous results for qubit
dephasing \cite{Krummheuer2002a,Doll} to spectral densities with arbitrary
$s=1,2,3,\ldots$

We will now briefly discuss the qualitatively different dynamics for 3D, 2D, and
1D substrates and present new analytical results, all based on
Eq.~(\ref{eqnCoherenceExact}), before comparing the error correction rates
$\omega_{\rm qec}$ for the three geometries.

\section{3D substrates}
Dephasing due to acoustic phonons in bulk geometries is described by
Eq.~\eqref{eqnCoherenceExact} for $s=3$ (for deformation-potential coupling, the
parameter $s$ equals the dimension of the substrate).
\begin{figure}[t]
\includegraphics{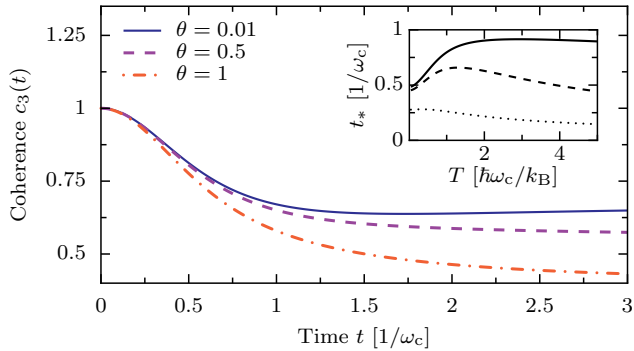} \caption{\label{figDynamicsBulk}(color
online). Time evolution of coherence for bulk structures with $s=3$ at different
scaled temperatures $\theta = k_\text B T / \hbar \omega_\text c$ with
$\alpha_{3} =0.1$. Inset: Non-monotonous temperature-dependence of the decay
time $t_\ast$ for different coupling strengths $\alpha_3 = 0.01$ (solid), $0.1$
(dashed), and $0.8$ (dotted). $t_\ast$ is defined as the time at which the
coherence has dropped `half-way' to $[1+c_{3}(\infty)]/2$,
see~Ref.~\cite{Vagov2004a}. }
\end{figure}
While a Markovian master equation for our model would predict no decoherence at
all, the exact dynamics in Fig.~\ref{figDynamicsBulk} exhibits a fast initial
decay of coherence, after which the coherence stabilizes to the final value
$c_3(\infty) = \exp[ - 8 \alpha_{3} (\theta^2 \Psi_1(\theta) - 1/2)]$, with
zero-temperature limit $\exp(-4\alpha_{3})$. Hence considerable initial decay
may occur even at low temperatures, as the solid (blue) curve in
Fig.~\ref{figDynamicsBulk} illustrates. Using the parameters of
Refs.~\cite{Krummheuer2002a,Krummheuer2005a} for GaAs/InAs self-assembled QDs we
obtain a typical value $\alpha_{3} = 0.8 \pm 0.3$ and $\omega_{c}= \unit[5 \cdot
10^{12}]{s^{-1}}$, in close agreement with the experimental
curves~\cite{Borri2001a,Vagov2004a}. For the exciton in the QD we find $l_3 =
10\,\unit{nm}$, a value in between the known confinement lengths of electron and
hole wave functions. In Ref.~\cite{Vagov2004a} a decay time  $t_{\ast}$ was
defined as the time at which half of the coherence that finally will get lost is
lost. Theory and experiment agreed well on the point that $t_{\ast}$ behaves
non-monotonously as a function of temperature. In the inset of
Fig.~\ref{figDynamicsBulk} we find for our model similar non-monotonous
behavior, another illustration that dephasing of real QDs is also well described
by our generic model.
Notice the short time scale of the decay, $t_{\ast}< \omega_{c}^{-1}$. There is
monotonous dependence on other parameters: $t_{\ast}$ gets longer by making the
QD larger, phonon velocity~$v_s$ slower, or the coupling~$\alpha_3$ smaller. As
Fig.~\ref{figDynamicsBulk} shows, the amount of coherence lost during short
times is considerable for this realistic choice of parameters. If for QIP
applications any phase errors beyond the percent level have to be corrected,
then this should occur well within~$t_{\ast}$.

Experimentally the coherence dynamics is usually inferred from absorption
spectra. The Fourier transform of the highly non-exponential coherence dynamics
in Fig.~\ref{figDynamicsBulk} predicts a highly non-Lorentzian spectrum. Since
$c_3(\infty)>0$, the zero-phonon line at frequency $\Delta/\hbar$ is even a delta
function, with weight $c_3(\infty)$ \cite{Duke1965a}. In practice this spectral
line has a very narrow finite width, finite due to slow processes not described
by the independent-boson model. However, with QIP in mind, we are more
interested in the fast initial decay, which shows up as the broad background of
the zero-phonon line.

\section{2D substrates}
Let us briefly also consider dephasing in planar geometries. To our knowledge,
experiments analogous to~\cite{Vagov2004a,Galland2008_short} with planar
geometries have not yet been carried out. The exact dynamics is obtained by
setting $s=2$ in Eq.~\eqref{eqnCoherenceExact}. Again a fast initial decay is
found, on a time scale $\omega^{-1}_\text c$. The coherence $c_{2}(t)$ does not
stabilize to a finite value, but vanishes algebraically $\propto 1/(\omega_c
t)^{8\alpha_{2}\theta}$, leading to a nontrivially broadened zero-phonon line in
the absorption spectrum.

\section{1D substrates}
For $s = 1$, the spectral density $J_{s}(\omega)$ becomes ohmic, {\em
i.e.}~linear in frequency up till the cutoff frequency $\omega_{c}$. This
describes dephasing due to acoustic phonons in 1D geometries such as nanotubes
and nanowires, for which it is well known that the coherence dynamics strongly
differs from the 3D case \cite{Galland2008_short,Lindwall2007}. Detailed
experiments have been performed only recently~\cite{Galland2008_short}. From
Eq.~\eqref{eqnCoherenceExact} we find the exact analytical expression
$\lambda_1(t) = 8\alpha_1 \{ \log \Gamma(\theta) -
\log |\Gamma(\theta[1+\im \omega_\text c t])| - \frac{1}{4}\log (1+\omega^2_\text c
t^2)\}$.
For long times $t \apprge \max\{\omega^{-1}_\text c, \hbar/k_\text B T\}$ this
gives  $c_{1}(t) = \kappa \exp(- t/T_{2})$, {\em i.e.}~exponential decay with
coherence time $T_2 = \hbar/4\pi\alpha_{1} k_\text B T$. This also gives the
zero-phonon line a finite width~\cite{Lindwall2007,Galland2008_short}. Master
equations would give the same $T_{2}$ time, but not the prefactor
$\kappa(\alpha_{1},\theta) = [2\pi\theta^{2\theta-1} /
\Gamma^2(\theta)]^{4\alpha_1}$    by which the initial non-exponential decay
remains noticeable also at long times. (See also the interesting discussion
in~Ref.~\cite{Wilhelm2005a_short}, where another $\kappa$ is found.) In the
limit $\theta \ll 1$ we find $\kappa = (2\pi \theta)^{4 \alpha_{1}}$. Dephasing
cannot be reduced indefinitely by lowering the temperature. Rather, the duration
of the non-exponential decay is increased. For the experiments in
Ref.~\cite{Galland2008_short} with isolated single-wall nanotubes we estimate
$\alpha_1 = 0.1\pm0.05$ and $\omega_{c} = \unit[20\cdot10^{12}]{s^{-1}}$. With
temperatures ranging from $5\unit{K}$ to $32\unit{K}$, {\em i.e.}~$\theta$
between $0.033$ and $0.21$, $\kappa$ assumes values between $0.49$ and $0.92$.
Thus non-exponential dephasing is important in state-of-the-art 1D systems, the
more so for lower temperatures.

\begin{figure}[t] \includegraphics{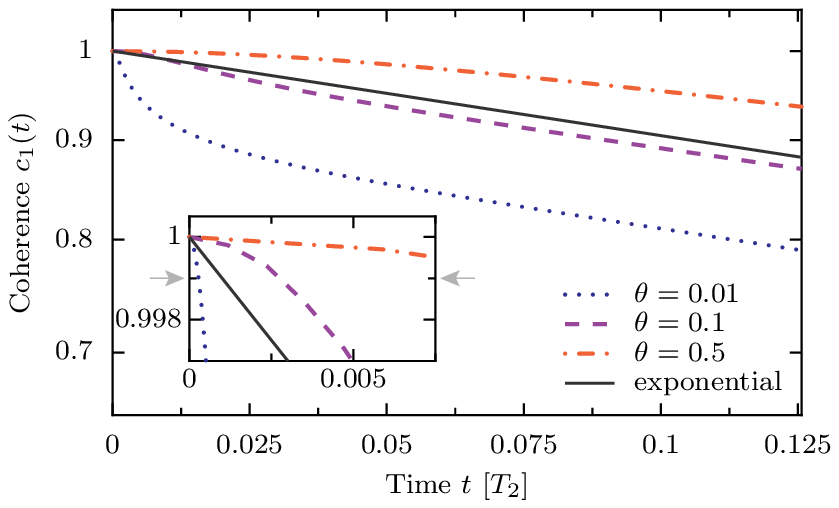}
\caption{\label{figDynamicsOhmic}(color online). Exact coherence dynamics for a
1D environment with ohmic spectral density with coupling strength $\alpha_1 =
0.01$, for several scaled temperatures $\theta = k_\text B T / \hbar
\omega_\text{c}$.
The time-axis is scaled by $\theta$. The dotted line shows the approximated
exponential decay as given by the $T_2$ time (see text). Inset: same as main
figure, but on a shorter time scale. The arrows indicate an error threshold
$\epsilon = 10^{-3}$.}
\end{figure}

Again, our main interest is the initial decay itself, rather than its effect at
long times. For $\theta \leq 1$, the time scale of the non-exponential decay  is
$\hbar/k_\text B T$, in clear contrast to the temperature-independent $t_{\ast}$
for bulk systems. Figure~\ref{figDynamicsOhmic}
shows the typical coherence dynamics of QDs on 1D substrates. We chose
$\alpha_{3}=0.01$, an order of magnitude smaller than in the
experiment~\cite{Galland2008_short}, but non-exponential decay would be
important even then.  The figure shows curves for three different temperatures,
as well as their master-equation approximations $\exp(-t/T_{2})$. The latter
coincide due to scaling of the time axis. By contrast, the three exact curves do
not coincide at all: the low-temperature curve ($\theta = 0.01$) is
systematically lower and the high-temperature curve higher than their
exponential-decay approximations. The inset of Fig.~\ref{figDynamicsOhmic} shows
that when asked how long $c_{1}(t)$ manages to stay above $0.999$, the
exponential-decay curve may be too optimistic, too pessimistic, or accurate by
chance.

\section{Rate of QEC}
There is the danger of comparing apples and oranges when studying the effect of
substrate dimensions. Fortunately, in state-of-the-art experiments both
$\alpha_{3}$ and $\alpha_{1}$ turn out to be of order $10^{-1}$ (see above). For
current and future experiments, it is therefore useful to depict the
$\alpha_s$-dependence of dephasing for all three geometries in the same range
$\alpha_s \simeq 0.1$ and smaller.

\begin{figure}[t]
\includegraphics{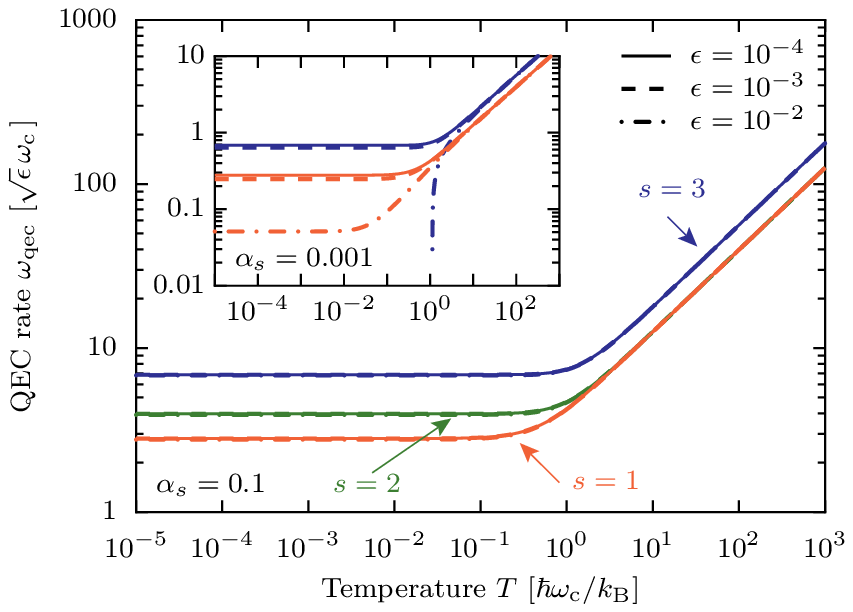}
\caption{\label{figQECRateVsTheta} (color online). Temperature dependence of the
quantum error correction rate $\omega_\mathrm{qec}$ for substrate geometries
with spectral densities $J_{s}(\omega)$ with different $s$ but equal couplings
$\alpha_s$, for several threshold values $\epsilon$. In the main figure
$\alpha_s = 0.1$, and $\alpha_s=0.001$ in the inset, where case $s=2$ is not
shown. The y-axes are scaled by $\sqrt{\epsilon}$. Curves are based on
Eq.~(\ref{eqnCoherenceExact}).}
\end{figure}

Let us now assume that phase errors have to be corrected if coherence drops from
$c_s(0) = 1$ to $c_s(2\pi/\omega_\text{qec}) = (1-\epsilon)$ for some error
threshold $\epsilon$, {\em i.e.}~from the exact dephasing dynamics
\eqref{eqnCoherenceExact} we identify the minimal rate $\omega_{\rm qec}$ at
which phase errors have to be corrected in order to preserve coherence in an
idle qubit. To be optimistic, we assume that in each step the error can be
corrected perfectly and instantaneously. Central idea is that structures with
lower rates $\omega_{\rm qec}$ are better suited for the implementation of QEC.
In Figure~\ref{figQECRateVsTheta}
we compare $\omega_{\rm qec}$ as a function of temperature for bulk, planar, and
linear substrate geometries. All couplings are $\alpha_{s}=0.1$. Notice that
$\omega_{\rm qec}$ on the vertical axis is scaled by $\sqrt{\epsilon}$. We find
that the curves for $\epsilon = 10^{-4}, 10^{-3}$, and $10^{-2}$ overlap for all
three geometries. Thus $\omega_{\rm qec}$ scales as $\sqrt{\epsilon}$, at least
for  current experimental couplings $\alpha_{s}=0.1$ and $\epsilon \le 10^{-2}$.
However, this simple scaling breaks down for very weak coupling $\alpha_{s}=
0.001$, as shown in the inset of Fig.~\ref{figQECRateVsTheta}. Scaling with
$\sqrt{\epsilon}$ only holds if the parabolic short-time approximation $c_{s}(t)
\approx 1 - \eta_{s} t^{2}$ is still valid at the time that $c_{s}(t)$ assumes
the threshold value $(1-\epsilon)$. We find $\eta_{s}(\alpha_s,\theta) =
2\alpha_s[2 (-\theta)^{s+1} \Psi_s(\theta) - s!]$, and for small $\epsilon$ the
error correction rate $\omega_{\text{qec}} = 2\pi\sqrt{\eta_s/\epsilon}$. For
$\alpha_{s} = 0.1$, the temperature dependence of $\eta_{s}$ and hence of
$\omega_{\rm qec}$ is completely negligible, even up to temperatures as high as
$\hbar \omega_{c}/k_{\rm B}$. This corroborates that the first stage of pure
dephasing is due to vacuum noise~\cite{Palma1996a,Doll}.
Fig.~\ref{figQECRateVsTheta} also shows that the rates for 1D are smaller than
for 3D or 2D geometries, although by  factors less than $10$. Generally we find
the central result that {\em linear substrate geometries like nanotubes and
nanowires will perform best}.
\begin{figure}
\includegraphics{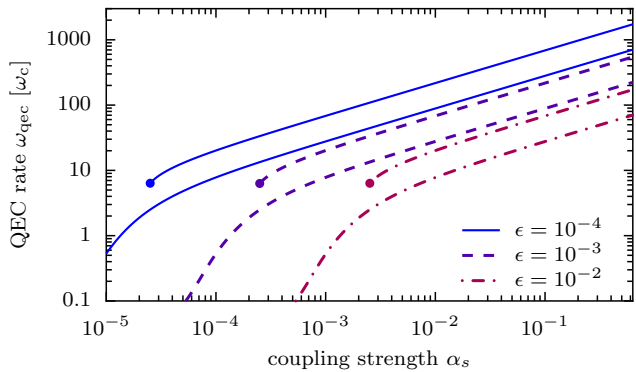}
\caption{\label{figQECRateVsAlpha}(color online). Quantum error correction rate
$\omega_\mathrm{qec}$ as a function of  coupling $\alpha_{s}$, for several error
thresholds $\epsilon$. Curves were made using Eq.~(\ref{eqnCoherenceExact}). For
all three pairs of curves, $s=1$ corresponds to the lower, and $s=3$ to the
upper curve. The curves for $s=3$ stop at critical coupling strengths indicated
by dots: For smaller $\alpha_{3}$, the coherence $c_{3}(t)$ never decays below
threshold. The temperature is $T =  0.01 \hbar \omega_\mathrm c/ k_\mathrm{B}$.}
\end{figure}
All rates $\omega_\mathrm{qec}$ in Fig.~\ref{figQECRateVsTheta} are high, in
between cutoff $\omega_\text c$ and qubit frequency $\Delta/\hbar \simeq
\unit[10^{15}]{s^{-1}}$, since already the fast initial decay reaches the error
threshold.

It would be very challenging to implement a quantum error correction protocol
for phase errors at such high rates.  In our model, lower rates $\omega_{\rm
qec}$ could be realized by reducing the cutoff frequency $\omega_{c}$ or the
couplings  $\alpha_{s}$. Figure~\ref{figQECRateVsAlpha} shows the effect of the
latter strategy. For the largest couplings $\alpha_{s} \simeq 0.1$, the message
is as in Fig.~\ref{figQECRateVsTheta}: given an error level $\epsilon$, linear
substrates have lower rates $\omega_\mathrm{qec}$ than planar or bulk
substrates. As the $\alpha_{s}$ are decreased, this message is essentially
unchanged, until suddenly for bulk substrates $\alpha_{3}$ gets so small that
$c_{3}(\infty)> 1-\epsilon$, {\em i.e.} the final coherence stabilizes above the
error threshold. In that situation  -- which can not occur for 1D substrates --
$\omega_{\rm qec}$ vanishes: no error correction is needed. State-of-the-art
exciton qubits in QDs have couplings that are more than one order of magnitude
larger than the largest critical coupling shown ({\em i.e.} for $\epsilon =
0.01$).
Obviously $\omega_{\rm qec}$  would also become smaller if larger errors
$\epsilon$ were allowed. The challenge is here to come up with QEC protocols
that tolerate larger faults.
All in all, Fig.~\ref{figQECRateVsAlpha} shows that for fixed
$\alpha_{s}=\alpha$ and $\omega_{c}$, linear substrate geometries are to be
preferred for their lower $\omega_{\rm qec}$, unless couplings can be
substantially reduced.

\section{Discussion and conclusions}
Inspired by recent measurements~\cite{Vagov2004a,Galland2008_short}, we have
analyzed the first few percents of loss of quantum coherence of a solid-state
exciton qubit on 3D, 2D, and 1D substrates. It is mainly this initial
decoherence that is important for QIP applications when supplemented with
fault-tolerant QEC.
We proposed and focused on the important quantity $\omega_{\rm qec}$, the
minimal rate at which quantum errors should be corrected. Its temperature
dependence turned out to be negligible. For QD exciton qubits in a bulk
substrate, the coherence may stabilize above the threshold, as  for spin
qubits~\cite{Mozyrsky2002a}, but corresponding couplings $\alpha_{3}$ are
currently not weak enough.

Let us return to our initial question: Is it  beneficial for QIP applications to
reduce the dimensions of the substrate, as far as dephasing is concerned? From a
master-equation perspective it is not, and worries about non-exponential decay
were expressed especially for 1D structures~\cite{Galland2008_short}. We have
presented analytical solutions of a generic but accurate independent-boson model
to show how fast initial dephasing occurs for 1D, 2D, and 3D geometries alike,
and identified the minimal rate at which single-qubit error have to be
corrected. In all cases the rates are high, in between the cutoff and the qubit
frequency, which poses a challenge for QIP applications. However, we found that
qubits on 1D substrates require the lowest error correction rates.

This work has been supported by the the DFG through SFB 631, by the German
Excellence Initiative via NIM, and by the Niels Bohr International Academy.


\end{document}